\documentclass[preprintnumbers,amsmath,11pt,amssymb,floatfix,superscriptaddress,nofootinbib]{article}
\usepackage{array}
\usepackage{relsize}
\usepackage{amsmath}
\usepackage{amssymb}
\usepackage{graphicx}
\usepackage[mathscr]{euscript}
\usepackage{tipa}
\usepackage{cite}
\usepackage{multicol}
\usepackage{caption}
\usepackage{subfigure}
\usepackage{lscape}
\usepackage{amssymb}
\usepackage[utf8]{inputenc}
\usepackage[english]{babel}
\usepackage{caption}

\usepackage[square,numbers]{natbib}
\graphicspath{{figures/}}
\def\mytitle#1{\setcounter{equation}{0}
\setcounter{footnote}{0}
\begin{flushleft}\Large\textbf{#1}\end{flushleft}
\vspace{0.25cm}}
\def\myname#1{\leftline{{\large #1}}\vspace{-0.13cm}}
\def\myplace#1#2{\small\begin{flushleft}\textit{#1}\\
\texttt{#2}\end{flushleft}}

\def\myclassification#1{\small\noindent
       #1\vspace{0.5cm}}
\usepackage{graphicx}

\DeclareFontFamily{OT1}{pzc}{}
\DeclareFontShape{OT1}{pzc}{m}{it}{<-> s * [0.900] pzcmi7t}{}
\DeclareMathAlphabet{\mathpzc}{OT1}{pzc}{m}{it}
\begin{document}

\mytitle{Viscous Dark Energy Accretion Activities : Sonic Speed, Angular Momentum and Mach Number Studies}

\myname{Sandip Dutta \footnote{duttasandip.mathematics@gmail.com}, Promila Biswas \footnote{promilabiswas8@gmail.com} and Ritabrata Biswas\footnote{biswas.ritabrata@gmail.com}}
\myplace{Department of Mathematics, The University of Burdwan, Golapbag Academic Complex, Burdwan -713104, Dist.:- Purba Barddhaman, State:- West Bengal, India.}{} 
 
\begin{abstract}
In this present article, we study different accretion properties regarding viscous accretion of dark energy. Modified Chaplygin gas is chosen as the dark energy candidate. Viscosity is encountered with the help of Shakura-Sunyaev viscosity parameter. We study sonic speed vs radial distance curves. We compare between adiabatic and dark energy dominated cases and follow that sonic speed falls as we go nearer to the central gravitating object. As viscosity is imposed, a threshold drop in accretion sonic speed is followed. Average rate of fall in accretion sonic speed is increased with black hole's spin. This is signifying that this kind of accretion is weakening the overall matter/energy infall. Specific angular momentum to Keplerian angular momentum ratio is found to fall as we go far from the black hole. Accretion Mach number turns high as we go towards the inner region and high wind Mach number is not allowed as we are going out. Combining, we conclude that the system weakens the feeding process of accretion.

\end{abstract}
\myclassification{}
\section{Introduction}
Just one Gyr after the Big Bang, super massive black holes (SMBHs) with masses of order $M_{BH} \sim 10^9 M_{\odot}$ started to exist. This is clear from studies of luminous quasars which are as early as $z\sim6-7$ \citep{DeRosa:2011}. To become such massive, they required continuous and exponential growth at the order of Eddington limit throughout the time universe has already experienced by then. For early SMBHs, these conditions might not be necessarily ubiquitous \citep{Volonteri:2016}. 

Mass accretion history of $\sim14000$ different haloes at $z=0$ using the ART code \citep{Benny:2017} in WMAP1 cosmology is identified by Wechsler et al \citep{Wechsler:2002}. The authors have found that the accretion histories of their present day haloes, on average, were well fitted by
\begin{equation}\label{eqn1}
M_{H}(z)=M_0 e^{-\alpha(z_{f})z}~~~~~~~~~~~~,
\end{equation}
where $M_{0}$ is the present day mass of a halo, $\alpha(z_{f})$ is a parameter which describes its formation epoch. For Einstein deSitter universe, the average mass accretion $\left< \dot{M}_H \right>$ onto a halo of mass $M_H$ has the same $M_H\dot{z}$ dependence like the equations 
\begin{equation}
\left< \dot{M}_H \right> \simeq M_H \bigg\rvert \frac{d\delta_c}{dt} \bigg\rvert f(M_H)~~~~~~~\&~~~~~~\frac{d\delta_c}{dt}=\frac{d\delta_c}{dD}\frac{dD}{dz}\frac{dz}{dt}~~~~~~~~~~,
\end{equation}
(where $\delta_c(t)$ = critical density above which an object will collapse to form a bound structure, $D(z)$ is linear growth factor and $f(M_H)$ = a weak function of halo mass) differing only in normalization $(\frac{d\delta_c}{dz}=1.686)$. With this idea, it is found to treat equation \eqref{eqn1} to be chosen as a sensible fit if we ignore slight mass dependencies of $\alpha(z_f(M_H))$ and $f(M_H)$ term.

Somerville \& Kolatt's Nbranch merger tree algorithm \citep{Somerville:1999} was used by Vanden Bosch (2002)\citep{Bosch:2002} to find a two parameter fit which better described the mass accretion history of concerned haloes. The relation provided was
$$\alpha=\left(\frac{z_f}{1.43}\right)^{-1.05}~~~~~~~~~~.$$
Though it is more common to present the definition of $z_f$ as an epoch where present day halo of interest had half of its present day mass
$$z_f=\frac{ln(2)}{\alpha}~~~~~~~~~~~~.$$
Mass accretion history of $\sim ~ 5,00,000$ haloes from the Millennium Simulation with $M_H > 10^{12}M_{\odot}$ and $0 \leq z \leq 6$ was investigated by Mc Bride et. al.\citep{McBride:2009} who found that approximately $25\%$ can be well described by equation \eqref{eqn1}. A second parameter $\beta$ has been introduced by them which requires another relation 
\begin{equation}\label{eqn2}
M_H(z) \propto (1+z)^{\beta} exp\{-\nu_z\}
\end{equation}
and this has provided a better fit to halo mass accretion history.

More recently, a joint dataset from the Millennium I and II simulations are used by Fakhori et al. \citep{Fakhouri:2009, Fakhouri:2010, Fakhouri:Michael:2010} to support equation \eqref{eqn2} to hold accross five decades in mass upto $z=15$. These works state very strongly that accretion rates on different haloes depend on the redshift zone in time it is embedded with. So different time in past we look back, different accretion rates might be possible to observe. Cosmic history of our universe is led by recent rapid progressions of precision cosmology. Simplest justifying model among many is $\Lambda$CDM which agrees with many cosmological observations \citep{Planck2015:AA}. Though this model suffers from different theoretical issues like fine tuning problem and cosmic coincidence problem \citep{Weinberg:1989, Zlatev:1999}.

Alternative dynamical dark energy models have been proposed \citep{COPELAND:2006, Bolotin:2015}. As the dark sector comprises of dark matter and dark energy, interaction between dark matter particles and dark energy is studied \cite{Wetterich:1994bg, Holden:2000, Carroll:1998, Damour:1990, Nusser:2004qu, Farrar:2004, Carroll:2009}. It is assumed that these two components stay coupled and hence we are motivated to study dark energy dominated accretion. As coupling with dark energy halo is considered, disc accretion model is preferred to be built.

Dark energy requires highly relativistic structure to fall gravitationally in. So we consider our disc around a black hole. First ever relativistic accretion was studied by F. Michel \cite{Michel:1972} where he has shown, by solving relativistic momentum and energy conservation equations, that the radiation speed gradient can be expressed as a ratio of two expressions. Denominator of the said expression turns zero when radial inward speed equals the sound speed through the fluid and to obtain a physical flow, we need to take help of L'Hospital's rule and a second degree first order differential equation will be constructed. This will give birth of two flows from the sonic point namely the accretion and the wind flows. Shakura and Sunyaev \cite{shakura_sunyaev_1973} added another milestone to the literature where shear viscosity, $\omega_{r\phi}$ is replaced by $\alpha_{ss}\rho c_s^2$, with $\alpha_{ss}$ is a parameter to regulate as per the model's viscosity requirement. 

 Accretion of dark energy was first studied by Babichev et al \cite{Babichev_2005}. They assumed some fluid permeated all over in the universe and responsible for late-time cosmic acceleration and the field equation followed by such fluid is enforcing it to weaken the corresponding accretion. Point to be noted, this article has considered spherical accretion. In between, we followed hierarchical clustering of dark matter is responsible for different structure creation \cite{Subramanian_2000}. Also, different works predicted interactions between dark matter and dark energy \cite{Farrar_2004, CAO_2013}. These support the presence of dark energy coupled with dark matter in different regions of galaxies. Does this ensure the presence of dark sectors in the vicinity of galactic cores? Very recently, the possibility of dark matter in the form of bosons are proposed to form self-gravitating bound structures in different galaxies \cite{PhysRevLett.121.151301}. Besides, authors of the reference \cite{Boshkayev_2019} compared the motion of test particles in the gravitational fields of supermassive black holes and dark matter core to find a noticeable discrepancy as the radial distance from the center is less than $100~AU$. In future, different finer observations like VLBI or Black Hole Cam Project might be able to more clear picture how much different the actual shadow of a black hole and the effect of a black hole mimicker is. As of now, we can not just exclude the theories that predict the existence of SMBH candidates like gravastars, boson stars etc. Dark energy and bulk viscosity can be formed out of delayed decay of dark matter \cite{PhysRevD.78.043525}. This, particularly, motivates us to examine the dark energy’s effect around a supermassive black hole. It is evident again accretion disc takes cylindrical structures \cite{First_M87}. 

First study of disc accretion of dark energy is found in 2011 \cite{Biswas_2011} where it was shown that if modified Chaplygin gas type agent is accreting upon a black hole, at a finite distance from the gravitating center,  it will increase the wind branch’s radial velocity so much that the speed will be equal to that of light. Physically, this interprets that matter is being thrown out with extreme speed and nothing is allowed to fall in. Viscosity is added to this model and studied in \cite{Dutta2017:SDRBEPJC2}, \cite{Dutta2019:Dutta}, \cite{Biswas_2019} and \cite{Roy:Biswas} to establish that viscosity will catalyze the whole process of accretion and as a result, the density of the accreting fluid drops drastically as we proceed towards the central black hole. Other physical quantities like Mach number, spin speed etc were not studied till date. However, the observations of the event horizon telescope enrich us a lot regarding these physical quantities. This motivates us towards the construction of this present article.

In this article, we are motivated to study viscous accretion onto a SMBH and to find its different consequences. In the next section, we will construct the mathematical problem. After that the solutions will be obtained and presented. Finally, we will conclude.
\section{Construction of the Model}
In the time interval 1980 to 2002, we observe a series of literatures trying to replace the general relativistic effect by pseudo Newtonian potentials. In this genre, Mukhopadhyay proposed one pseudo Newtonian force as \citep{Mukhopadhyay:2002} 
\begin{equation}
\mathscr{F_G}\mathpzc{(x,J)}=\mathpzc{\frac{(x^2-2J\sqrt{x}+J^2)^2}{x^{3}\{\sqrt{x}(x-2)+J\}^2}}~~~~~~~~~~,
\end{equation}
where $\mathpzc{x}=\left(\frac{\mathpzc{R}}{\frac{GM}{c^2}}\right)$, $\mathpzc{J}=\left(\frac{\mathpzc{A}}{c}\right)$, $\mathpzc{R}$, $G$, $M$, $c$ and $\mathpzc{A}$ are radial distance from the central object, Newton's gravitational constant, mass of the central gravitating object, speed of light and Kerr spinning parameter respectively. 

 Axisymmetric accretion study was constructed by Novikov and Thorne \cite{Novikov:1973kta} in 1973. Quasi spherical model \cite{Fabian, Stoeger:Nature, Stoeger}, however,  is treated to be the process which is involved in a large number of high energy astrophysical situations. These models support both galactic and extragalactic compact X-ray sources, novae, dwarf novae, active galactic nuclei, globular cluster X-ray sources, X-ray and $\gamma$-ray bursts and quasars \cite{Stoeger, Igumenshchev}. Chakrabarti studied transonic properties of conical shaped accretion flows, known as Wedge-Shaped flow \cite{Chakrabarti}.

Accretion model keeping hydrostatic equilibrium in vertical direction is also found in literature. To keep simplicity, necessity to construct models having all the salient features of the original problem should be given priority. Cylindrical model, conical model and flow in vertical equilibrium model have identical physical properties, in spite of the fact that they are constructed based on fundamentally different assumptions. Results of one model can be  obtained by changing the physical parameters of the other \cite{Chakrabarti:2001}. Alongside, the reference \cite{Sankhasubhra} and the references therein are also the relevant literature. 

In our article, we assume axisymmetric flow and radiative losses of internal energy. Hence the disc turns geometrically thin. In this case, accretion occurs owing to the overflow of the effective potential barrier near the black hole. This is equivalent to the case of the Roche lobe overflowing star present in a binary. Viscosity considered is due to the small scale turbulence and described by $\alpha_{SS}$-prescription. In Kerr space-time, vertical hydrodynamical equilibrium of stationary thin and slim accretion discs is also studied in \cite{Abramowicz_1997}. Cylindrical polar coordinates are best to explain axisymmetric models. Hence we consider cylindrical polar coordinates and construct the model. So at a distance $\mathpzc{x}$, quantity of matter inflow should be $\dot{\mathscr{M}}=-4\pi\mathpzc{xU}\rho\mathpzc{h(x)}$ and if this rate is taken as constant, i.e., equation of continuity leads us to
\begin{equation}
\frac{d}{d\mathpzc{x}}\bigg\{\mathpzc{xU}\rho\mathpzc{h(x)}\bigg\}=0~~~~~~~~~~~,
\end{equation}
here $\mathpzc{U}=\frac{\mathpzc{V_r}}{c}$, $\mathpzc{V_r}$ is radial inward speed and $\rho$ (accretion disc height) presents the vertically averaged density as we are not considering changes in density or any other parameters with height.

Next we will consider Navier Stokes equation for steady state given as
\begin{equation}
\rho(\overrightarrow{\mathpzc{U}}.\overrightarrow{\nabla})\overrightarrow{\mathpzc{U}}=-\overrightarrow{\nabla}p+\rho \mathpzc{\nu} {\nabla}^2 \overrightarrow{\mathpzc{U}}-\overrightarrow{\mathscr{F_G}\mathpzc{(x,J)}}
\end{equation}
and the radial momentum balance equation turns out as
\begin{equation}
\mathpzc{U}\frac{d\mathpzc{U}}{d \mathpzc{x}}+\frac{1}{\rho}\frac{d p}{d \mathpzc{x}}-\frac{\lambda^2}{\mathpzc{x}^3}+\mathscr{F_G}\mathpzc{(x,J)}=0~~~~~~~~~~~,
\end{equation}
where $p$ is the accretion fluid pressure, $\mathpzc{\nu}$ is the viscous coefficient and $\lambda$ is specific angular momentum. 

Again we get the azimuthal momentum balance equation as
\begin{equation}
\mathpzc{U}\frac{d\lambda}{d\mathpzc{x}}=\frac{1}{\mathpzc{x}\rho \mathpzc{h(x)}}\frac{d}{d\mathpzc{x}}\left[{\mathpzc{x}^2} \alpha_{ss} (p+\rho {\mathpzc{U}^2}) \mathpzc{h(x)} \right]~~~~~~~~.
\end{equation}

We will assume that the vertical equilibrium and be able to write the vertical momentum balance equation as $\mathpzc{h(x)}=c_s \sqrt{\frac{\mathpzc{x}}{\mathscr{F_G}\mathpzc{(x,J)}}}.$

As we have stated earlier, we will choose dark energy as accreting agent. We take modified Chaplygin gas as the candidate of dark energy. This model of dark energy was proposed within the framework of FLRW cosmology \cite{Arun_2017}. The equation of state is given by
$p=\alpha \rho-\frac{\beta}{\rho^n}$, where $\alpha$, $\beta$ and $\mathpzc{U}$ are free parameters.

This equation of state is able to interpolate from matter dominated era to the cosmological constant dominated era. Xu and Lu have constrained the parameter $n$ and related cosmological parameters like $\Omega_b{h^2}$, $H_0$ as
$$n= 0.033^{+0.066}_{-0.071}(1\sigma)^{+0.096}_{-0.087}(2\sigma)~~~~~~~~~~~~~,$$
$$\Omega_b h^2=0.0233^{+0.0023}_{-0.0016}(1\sigma)^{+0.0029}_{-0.0020}(2\sigma)~~~~~~~~~~~and$$
$$H_0=69.97^{+2.87}_{-2.78}(1\sigma)^{+3.48}_{-3.08}(2\sigma)~~~~~~~~~~~~,$$
with a minimum chi squared $=519.342$. 

From the equation of state of modified Chaplygin gas, we obtain the sound speed squared as 
\begin{equation}
{c_s}^2=\frac{\partial p}{\partial \rho} = \alpha + \frac{\beta n}{\rho^{n+1}}~~~~~~~.
\end{equation}
With the help of the continuity equation, we obtain 
\begin{equation}\label{eqn3}
\frac{1}{\rho}\frac{d p}{d \mathpzc{x}} = -\frac{2{c_s}^3}{(n+1)({c_s}^2-\alpha)} \frac{dc_s}{d \mathpzc{x}}=-\frac{1}{n+1}\frac{d{c_s}^2}{d \mathpzc{x}}-\frac{\alpha}{n+1}\frac{d}{d \mathpzc{x}}\{ln({c_s}^2-\alpha)\}
\end{equation}
and the gradient of sound speed is obtained as 
\begin{equation}\label{eqn4}
\frac{dc_s}{d\mathpzc{x}}= \bigg\{\frac{3}{2\mathpzc{x}} -\frac{1}{2 \mathscr{F_G}\mathpzc{(x,J)}} \frac{d \mathscr{F_G}\mathpzc{(x,J)}}{d \mathpzc{x}} + \frac{1}{\mathpzc{U}} \frac{d \mathpzc{U}}{d \mathpzc{x}} \bigg\}\left\{\frac{(n+1)c_s ({c_s}^2 - \alpha)}{(1-n){c_s}^2+\alpha(n+1)} \right\}~~~~~~~~.
\end{equation}
Azimuthal momentum balance equation turns to be
$$\frac{d \lambda}{d \mathpzc{x}} = \frac{\mathpzc{x}\alpha_{ss}}{\mathpzc{U}}\left[\frac{1}{2}\bigg\{\frac{5}{\mathpzc{x}}-\frac{1}{\mathscr{F_G}\mathpzc{(x,J)}} \frac{d \mathscr{F_G}\mathpzc{(x,J)}}{d \mathpzc{x}}\bigg\}\bigg\{\frac{(n+1)\alpha-{c_s}^2}{n}+\mathpzc{U^2}  \bigg\}\right.$$
\begin{equation}\label{eqn5}
\left. +2\mathpzc{U}\frac{d\mathpzc{U}}{d\mathpzc{x}}+\left\{\left(\frac{(n+1)\alpha-{c_s}^2}{n}+\mathpzc{U^2} \right)\frac{1}{c_s}-({c_s}^2+\mathpzc{U^2})\left(\frac{1}{n+1}-\frac{2c_s}{{c_s}^2-\alpha}\right) \right\}\frac{dc_s}{d\mathpzc{x}}\right]
\end{equation}
and replacing $\frac{1}{\rho} \frac{d p}{d \mathpzc{x}}$ in \eqref{eqn3}, we obtain,
\begin{equation}\label{eqn6}
\frac{d\mathpzc{U}}{d\mathpzc{x}} = \frac{\frac{\lambda^2}{\mathpzc{x^3}}-\mathscr{F_G}\mathpzc{(x,J)}+\left\{\frac{3}{\mathpzc{x}}-\frac{1}{\mathscr{F_G}\mathpzc{(x,J)}}\frac{d \mathscr{F_G}\mathpzc{(x,J)}}{d \mathpzc{x}} \right\}\frac{{c_s}^4}{\{(1-n){c_s}^2+\alpha(n+1)\}}}{\mathpzc{U}-\frac{2{c_s}^4}{\mathpzc{U}\{(1-n){c_s}^2+\alpha(n+1)\}}}~~~~~~~~~~.
\end{equation}

After we have completed the constructions of the required differential equations \eqref{eqn4},\eqref{eqn5} and \eqref{eqn6}, we will require to set the initial conditions. To do so, we will take help of a basic crucial nature of accretion around black holes. Unlike neutron stars and other accretors, black holes do not possess solid  surfaces. Near the event horizon, the accretion speed becomes almost equal to speed of light. Besides it is obvious that very far from the accretion disc, where a particle just enters into the disc due to the angular momentum transportation process, the radial inward speed is very low, almost equal to zero. Now throughout the accretion disc, the radial speed starts from zero and reaches $1$ where $c$ is taken to be unit. In between at some $\mathpzc{x}={\mathpzc{x}}_c$ the speed should be equal to $\frac{\sqrt{2}{c_s}^2}{\{(1-n){c_s}^2+\alpha(n+1)\}^{\frac{1}{2}}}$ and the denominator will vanish, i.e., $D({\mathpzc{U}}_c,c_{sc})=0$. Now to have a continuous physical flow, the numerator should vanish as well, $N({\mathpzc{x}}_c,~\mathpzc{J},~c_{sc})=0$ which provides the value of sound speed at ${\mathpzc{x}}_c$. Using this, we obtain ${\mathpzc{U}}_c$ at ${\mathpzc{x}}_c$ from $D({\mathpzc{U}}_c,c_{sc})=0$. $\lambda_c($ at $\mathpzc{x} ={\mathpzc{x}}_c)$ is assumed such that a physical flow is obtained. 

Now applying the L'Hospital's rule  in equation (\ref{eqn6}) at the critical point (say $\mathpzc{x} = \mathpzc{x}_c$) we get a quadratic equation of $\frac{d\mathpzc{U}}{d\mathpzc{x}}$ in the form,
\begin{equation}\label{quadratic}
A \left( {\frac{d\mathpzc{U}}{d\mathpzc{x}}}\right)_{\mathpzc{x} = \mathpzc{x}_c}^2 + B \left( \frac{d\mathpzc{U}}{d\mathpzc{x}} \right)_{\mathpzc{x} = \mathpzc{x}_c} + C =0.
\end{equation}
Where \\
$A= 1 + \frac{1}{c_{sc}^2} - \frac{4 \left( n+1 \right) \left( c_{sc}^2 -\alpha \right)}{{\left\lbrace \left( 1-n \right) c_{sc}^2 +\alpha \left( n+1 \right) \right\rbrace}^2}$,\\
$B= \frac{4 \lambda \mathpzc{U}}{\mathpzc{x}_c^3} + \frac{2 \mathpzc{U}_c c_{sc} \left( 1-n \right)}{ \left\lbrace \left( 1-n \right) c_{sc}^2 +\alpha \left( n+1 \right) \right\rbrace} - \frac{2 \mathpzc{U}_c \left( n+1 \right) \left( c_{sc}^2 -\alpha \right)}{{\left\lbrace \left( 1-n \right) c_{sc}^2 +\alpha \left( n+1 \right) \right\rbrace}^2} \left\{ \frac{3}{\mathpzc{x}_c} - \frac{1}{\mathscr{F_G}\mathpzc{(x,J)}} \left(\frac{d\mathscr{F_G}\mathpzc{(x,J)}}{d\mathpzc{x}}\right)_{\mathpzc{x} = \mathpzc{x}_c} \right\} + \\
\left[ \left( c_{sc}^2 + \mathpzc{U}_c^2 \right) \left( \frac{1}{n+1} \frac{2 c_{sc}}{c_{sc}^2 - \alpha} \right)-\left\{ \frac{3}{\mathpzc{x}_c} - \frac{1}{\mathscr{F_G}\mathpzc{(x,J)}} \left(\frac{d\mathscr{F_G}\mathpzc{(x,J)}}{d\mathpzc{x}}\right)_{\mathpzc{x} = \mathpzc{x}_c} \right\} \frac{4 \mathpzc{U}_c }{c_{sc}{\left\lbrace \left( 1-n \right) c_{sc}^2 +\alpha \left( n+1 \right) \right\rbrace}}-  \left( \frac{\left(n+1 \right) \alpha - c_{sc}^2}{n} + {\mathpzc{U}}^2 \right)  \frac{1}{c_{sc}} \right] \left\lbrace \frac{ \mathpzc{U}_c \left( n+1 \right) \left( c_{sc}^2 -\alpha \right)}{2 c_{sc}^3} \right\rbrace $\\
and $C= D+ E+ F$.\\
The values of $D$, $E$ and $F$ are,\\
$D= \left[ \left( c_{sc}^2 +\mathpzc{U}_c^2 \right) \left( \frac{1}{n+1} \frac{2 c_{sc}}{c_{sc}^2 - \alpha} \right)-\left\{ \frac{3}{\mathpzc{x}_c} - \frac{1}{\mathscr{F_G}\mathpzc{(x,J)}} \left(\frac{d\mathscr{F_G}\mathpzc{(x,J)}}{d\mathpzc{x}}\right)_{\mathpzc{x} = \mathpzc{x}_c} \right\} \frac{4 \mathpzc{U}_c }{c_{sc}{\left\lbrace \left( 1-n \right) c_{sc}^2 +\alpha \left( n+1 \right) \right\rbrace}}-  \left( \frac{\left(n+1 \right) \alpha - c_{sc}^2}{n} + {\mathpzc{U}}^2 \right)  \frac{1}{c_{sc}} \right] \\
\left\{ \frac{3}{2\mathpzc{x}_c} -\frac{1}{2\mathscr{F_G}\mathpzc{(x,J)}} \left(\frac{d\mathscr{F_G}\mathpzc{(x,J)}}{d\mathpzc{x}} \right)_{\mathpzc{x} = \mathpzc{x}_c} \right\} \left\lbrace \frac{\left( n+1 \right) c_{sc} \left( c_{sc}^2 -\alpha \right)}{\left( 1-n \right) c_{sc}^2 + \alpha \left( n+1 \right)} \right\rbrace $,\\
$E= \left\{ \frac{3}{\mathpzc{x}_c} - \frac{1}{\mathscr{F_G}\mathpzc{(x,J)}} \left(\frac{d\mathscr{F_G}\mathpzc{(x,J)}}{dx}\right)_{\mathpzc{x} = \mathpzc{x}_c} \right\} \frac{\mathpzc{U}_c c_{sc} \left( 1-n \right)}{{\left\lbrace \left( 1-n \right) c_{sc}^2 +\alpha \left( n+1 \right) \right\rbrace}} + \left(\frac{d\mathscr{F_G}\mathpzc{(x,J)}}{d\mathpzc{x}}\right)_{\mathpzc{x} = \mathpzc{x}_c}$  
$$+\left\lbrace \frac{1}{{\mathscr{F_G}\mathpzc{(x,J)}}^2} \left(\frac{d\mathscr{F_G}\mathpzc{(x,J)}}{d\mathpzc{x}} \right)^2_{\mathpzc{x} = \mathpzc{x}_c}-\frac{1}{\mathscr{F_G}\mathpzc{(x,J)}} \left(\frac{d^2 \mathscr{F_G}\mathpzc{(x,J)}}{d\mathpzc{x}^2}\right)_{\mathpzc{x} = \mathpzc{x}_c}-\frac{3}{\mathpzc{x}_c^2}\right\rbrace \frac{\mathpzc{U}_c^2}{2}~~~~~~~~~~,$$\\
$F= \frac{\lambda \alpha_{ss}}{\mathpzc{x}_c^2 \mathpzc{U}_c} \left\{ \frac{1}{\mathscr{F_G}\mathpzc{(x,J)}} \left(\frac{d\mathscr{F_G}\mathpzc{(x,J)}}{d\mathpzc{x}}\right)_{\mathpzc{x} = \mathpzc{x}_c} - \frac{5}{\mathpzc{x}_c} \right\} \left\lbrace \frac{\left(n+1 \right) \alpha - c_{sc}^2}{n} + \mathpzc{U}_c^2 \right\rbrace  $.

While we are working with a particular type of EoS we supply $n$ and $\alpha$, i.e., vanishing denominator, $D\left( \mathpzc{U}_c, c_{sc}, n, \alpha\right)=0$ will be reduced to an equation $D\left( \mathpzc{U}_c, c_{sc}\right)=0$, which tells us the relation between $\mathpzc{U}_c$ and $c_{sc}$, on the other hand vanishing numerator, i.e., $N\left( \lambda_c, \mathpzc{x}_c, \mathpzc{J}, c_{sc}, n, \alpha\right)=0$ will be reduced to $N\left( \lambda_c, \mathpzc{x}_c, c_{sc}\right)=0$ once we provide the information about the accreting fluid and the rotation of the black hole. We will choose where the sonic point will be formed and adjust the value of angular momentum at the critical point which ultimately form an algebraic equation of $c_{sc}$,  $N \left( c_{sc}=0 \right)$ From there we can easily solve $c_{sc}$ and hence calculate the value of $\mathpzc{U}_c$. These will be used as the initial values to solve (\ref{quadratic}).

\section{Solutions and Graphical Interpretations}
\begin{figure}
	\begin{center}
		Fig 1.1(a)\\
	\end{center}
	\includegraphics[height=10cm,width=16cm]{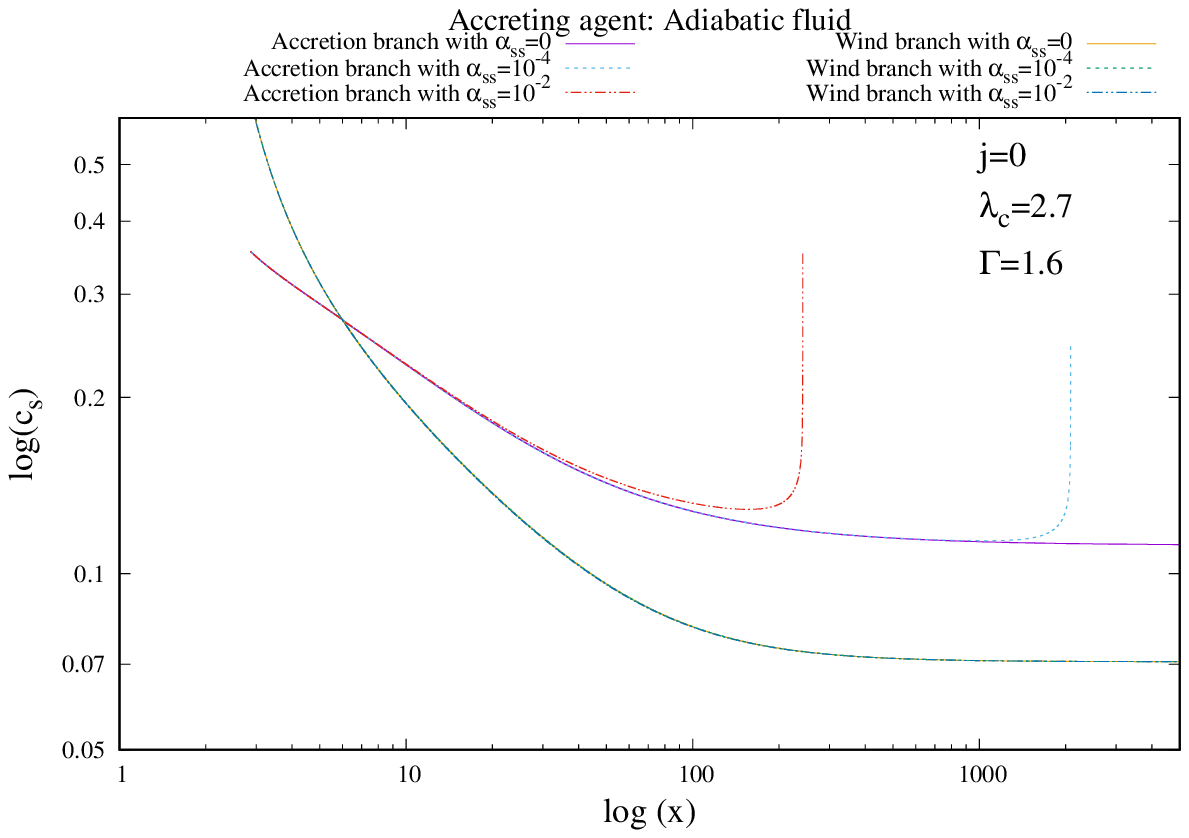}
	\begin{center}
		Fig 1.1(b)\\
	\end{center}
	\includegraphics[height=10cm,width=16cm]{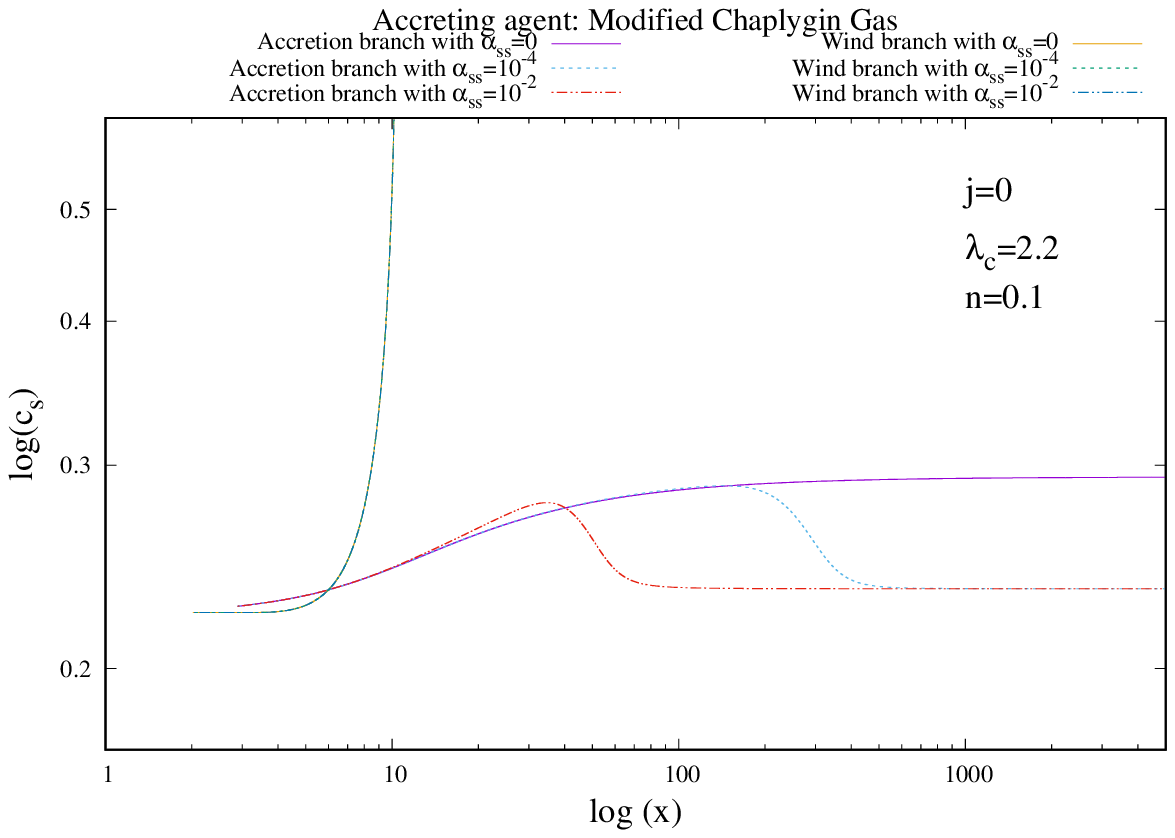}\\
	{Figure 1.1 : Plots of accretion and wind branches for $log(c_s)$ (log of sound speed) vs $log(\mathpzc{x})$ for non-rotating black hole with different viscosity parameter (showed in different colours). (a) For adiabatic fluid accretion (b) For MCG accretion.}
\end{figure}

\begin{figure}
	\begin{center}
		Fig 1.2(a)\\
	\end{center}
	\includegraphics[height=10cm,width=16cm]{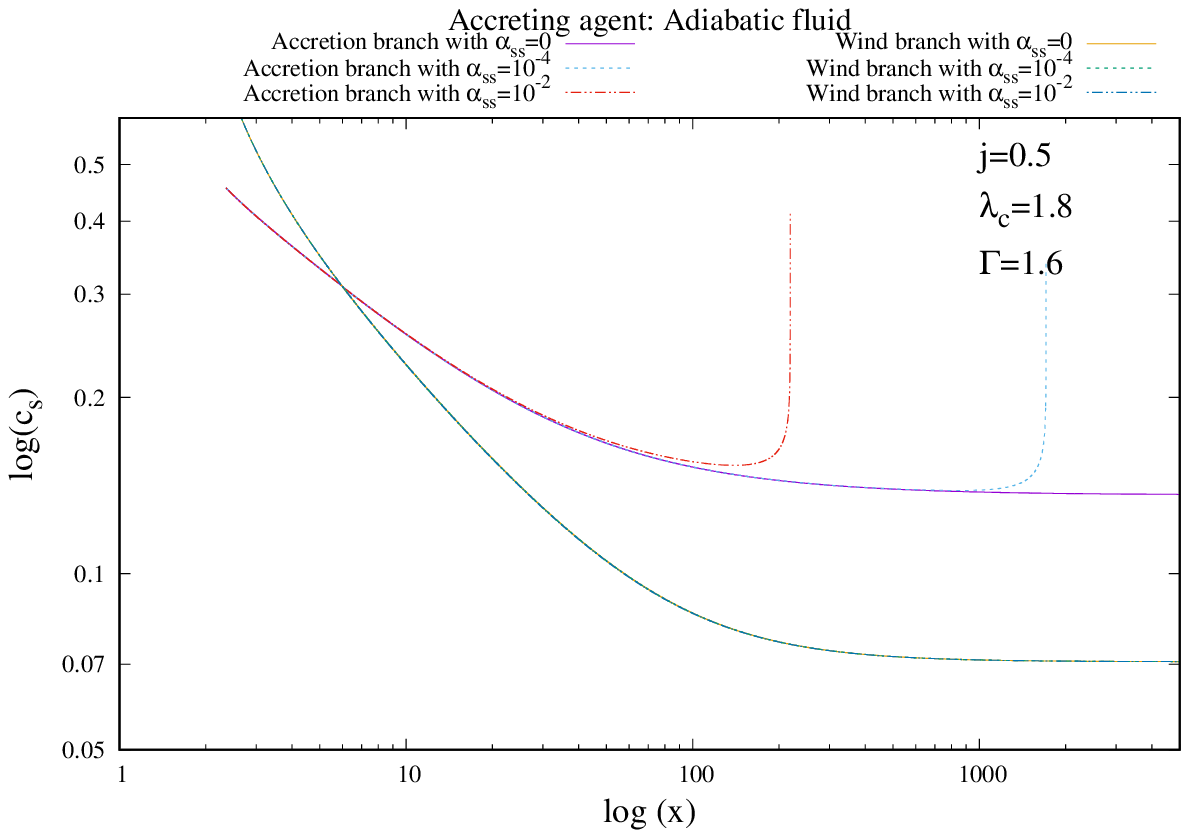}
	\begin{center}
		Fig 1.2(b)\\
	\end{center}
	\includegraphics[height=10cm,width=16cm]{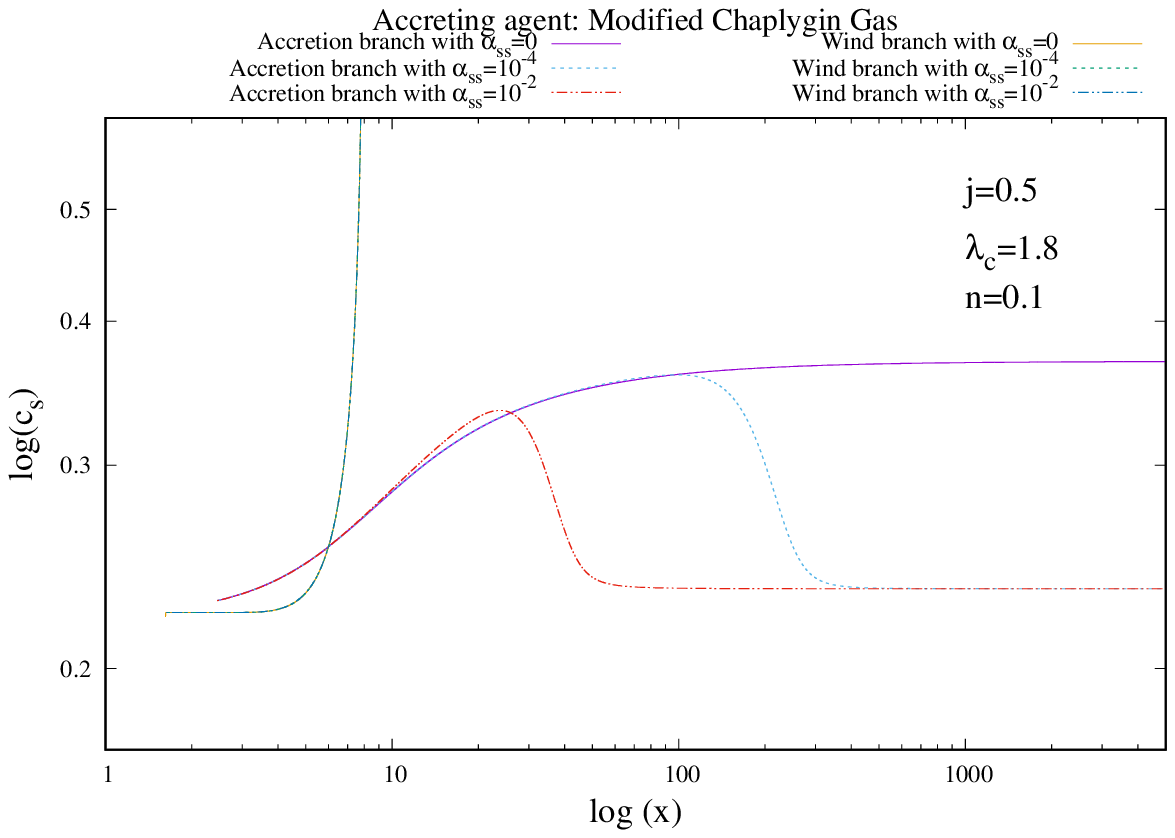}\\
	{Figure 1.2 : Plots of accretion and wind branches for $log(c_s)$ (log of sound speed) vs $log(\mathpzc{x})$ for rotating black hole ($spin~ parameter~\mathpzc{J}=0.5$) with different viscosity parameter (showed in different colours). (a) For adiabatic fluid accretion (b) For MCG accretion.}
\end{figure}

\begin{figure}
	\begin{center}
		Fig 1.3(a)\\
	\end{center}
	\includegraphics[height=10cm,width=16cm]{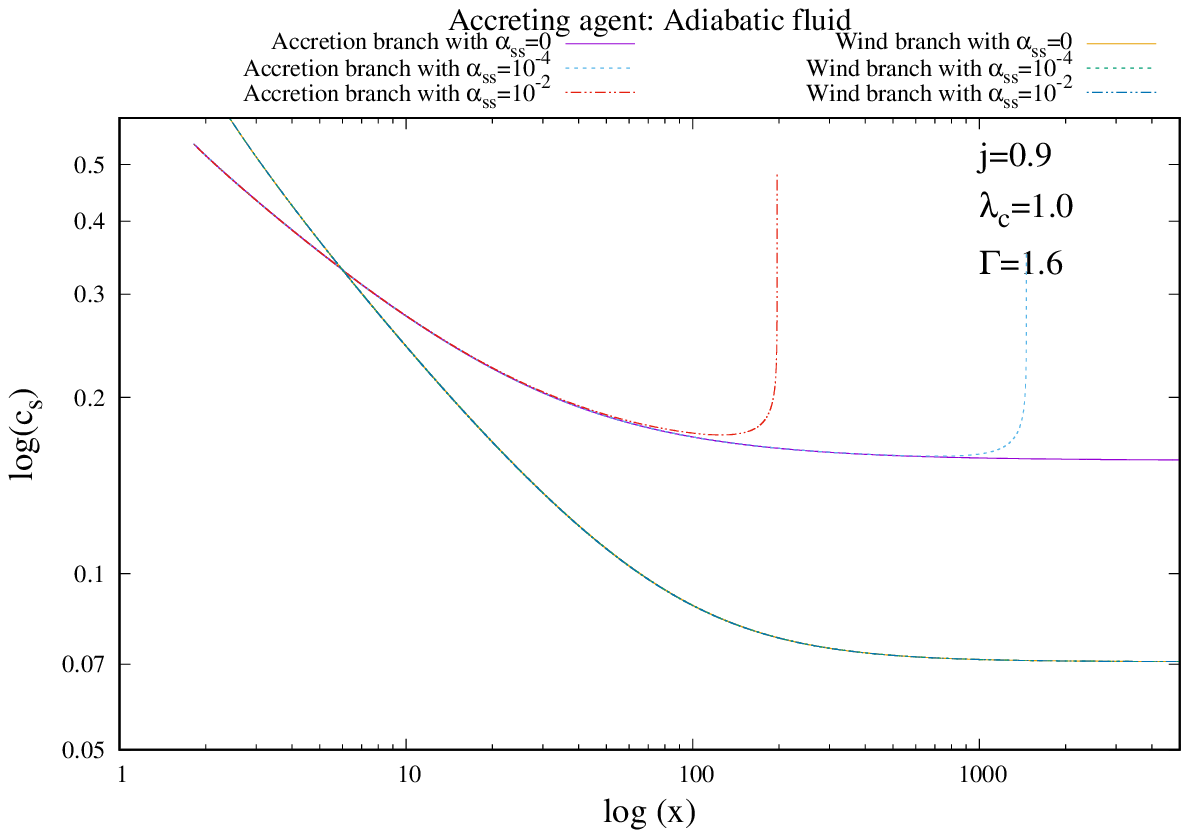}
	\begin{center}
		Fig 1.3(b)\\
	\end{center}
	\includegraphics[height=10cm,width=16cm]{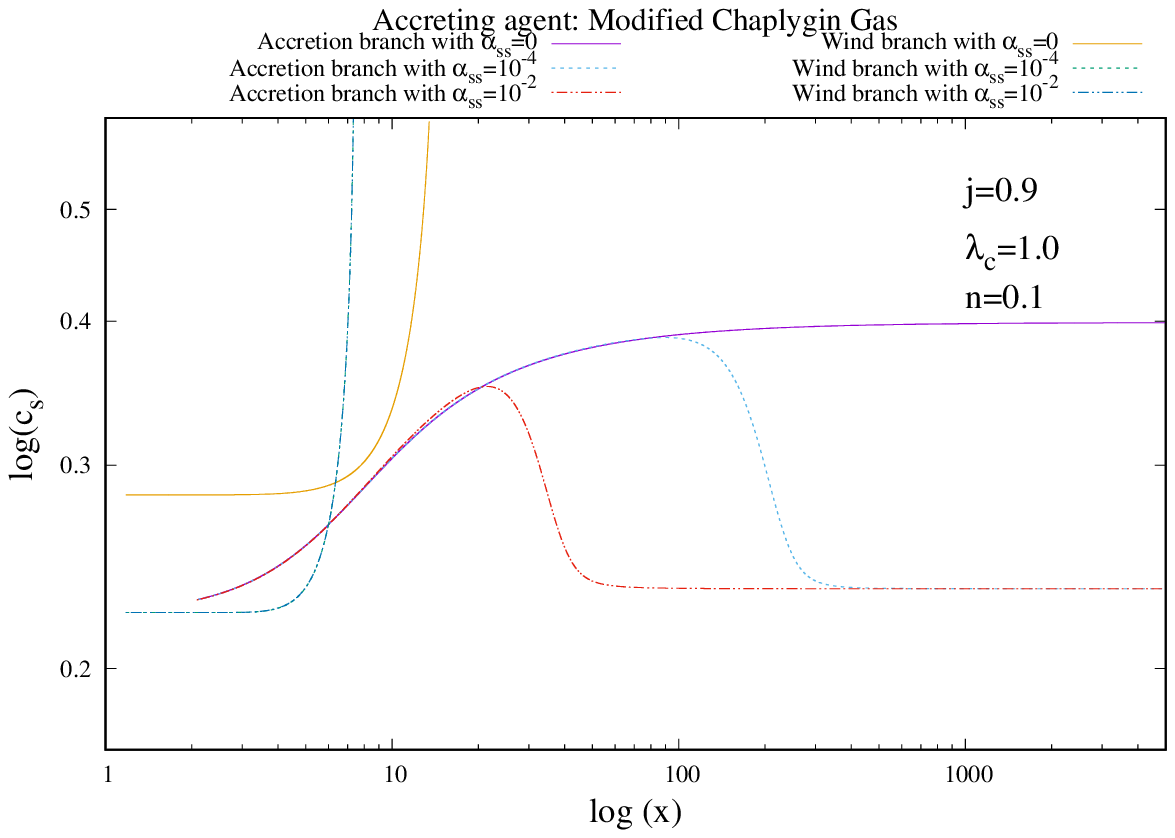}\\
	{Figure 1.3 : Plots of accretion and wind branches for $log(c_s)$ (log of sound speed) vs $log(\mathpzc{x})$ for rotating black hole ($spin~ parameter~\mathpzc{J}=0.9$) with different viscosity parameter (showed in different colours). (a) For adiabatic fluid accretion (b) For MCG accretion.}
\end{figure}
\begin{figure}
	\begin{center}
		Fig 2.1(a)\\
	\end{center}
	\includegraphics[height=10cm,width=16cm]{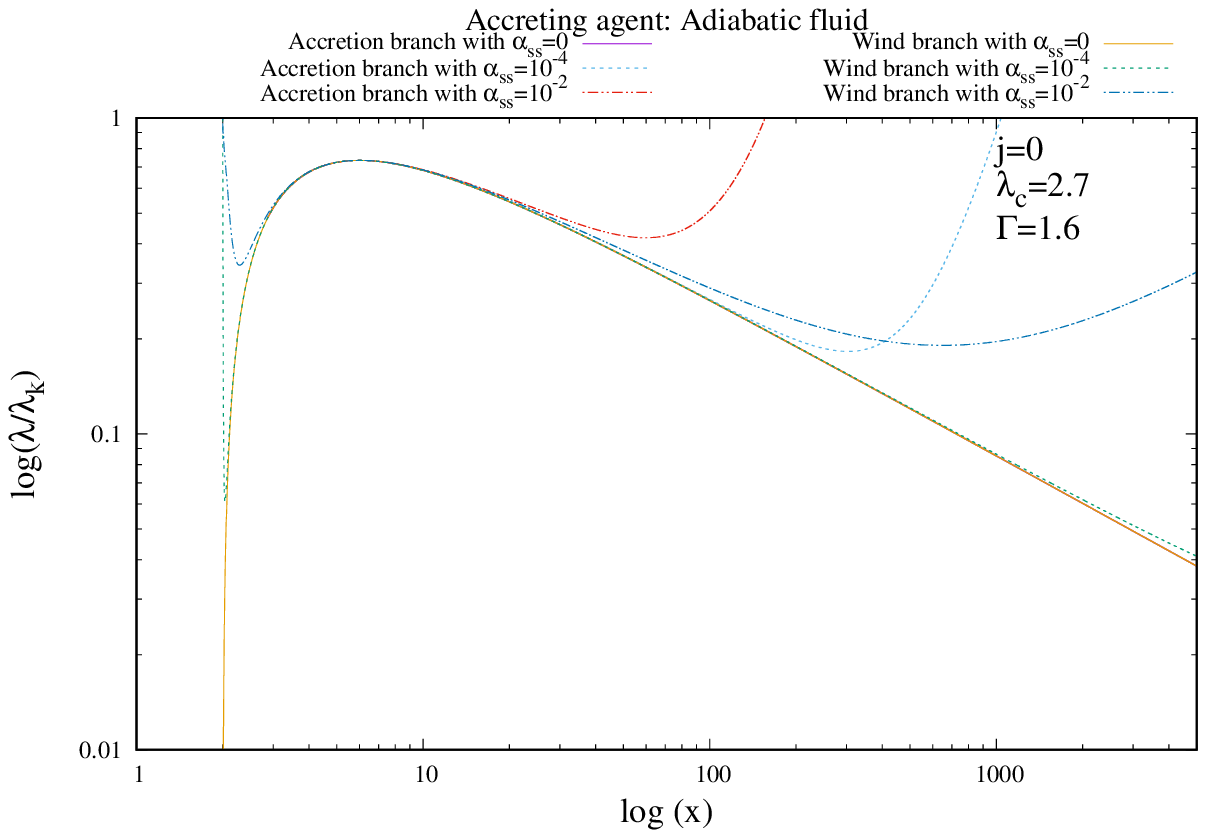}
	\begin{center}
		Fig 2.1(b)\\
	\end{center}
	\includegraphics[height=10cm,width=16cm]{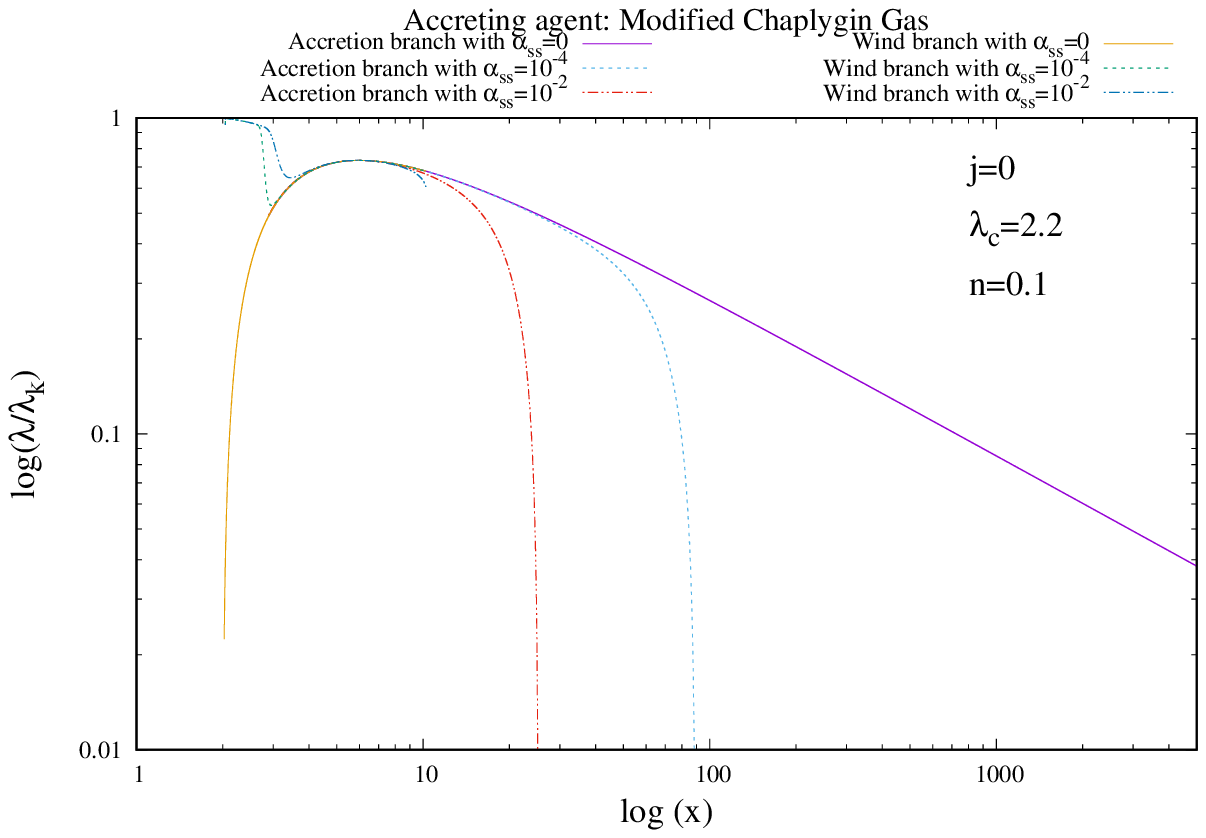}\\
	{Figure 2.1 : Plots of accretion and wind branches for $log(\frac{\lambda}{\lambda_k})$ vs $log(\mathpzc{x})$ for non-rotating black hole with different viscosity parameter (showed in different colours). (a) For adiabatic fluid accretion (b) For MCG accretion.}
\end{figure}

\begin{figure}
	\begin{center}
		Fig 2.2(a)\\
	\end{center}
	\includegraphics[height=10cm,width=16cm]{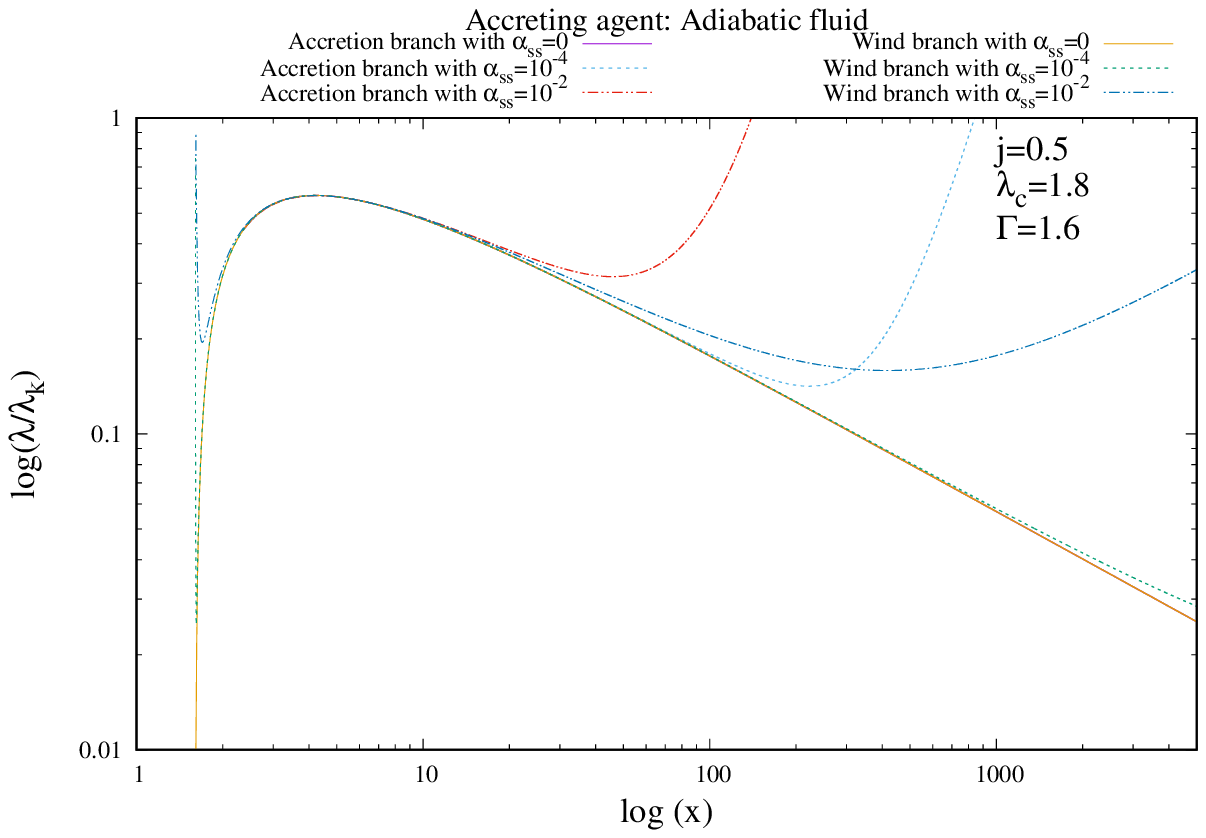}
	\begin{center}
		Fig 2.2(b)\\
	\end{center}
	\includegraphics[height=10cm,width=16cm]{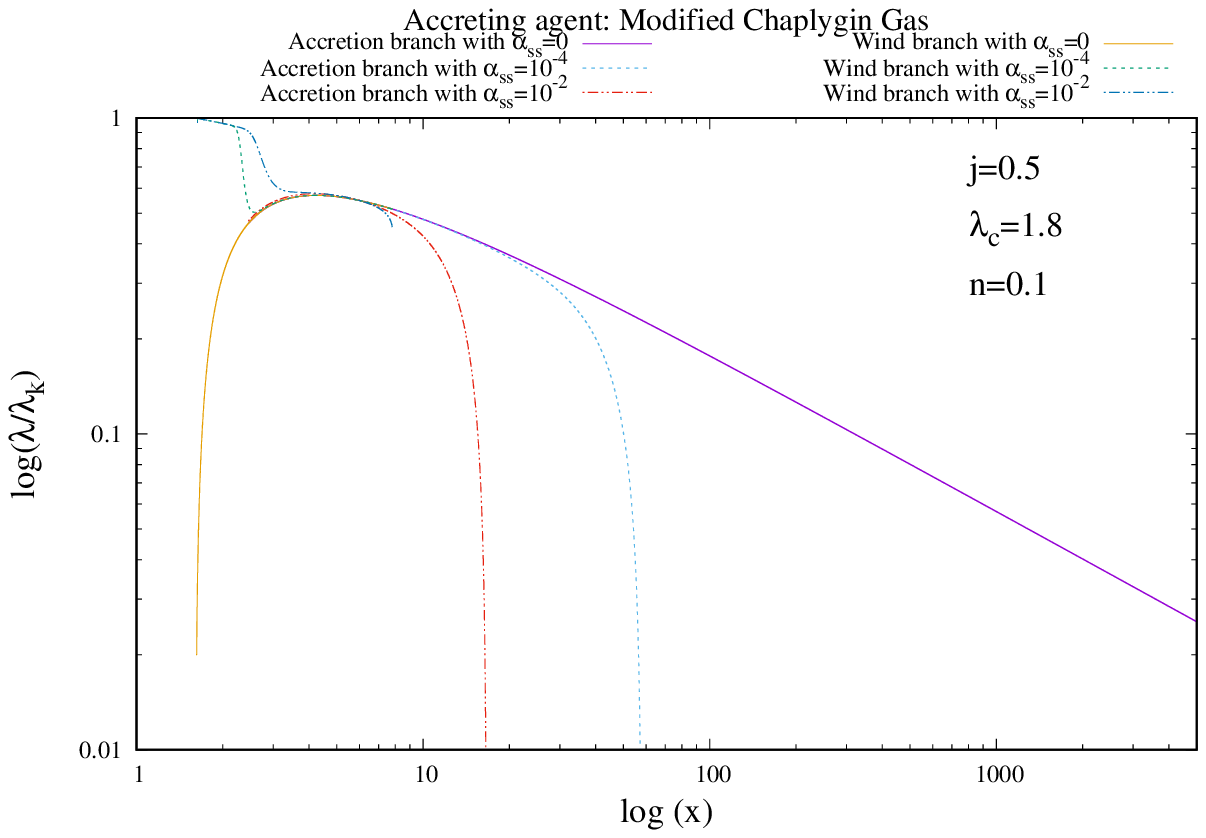}\\
	{Figure 2.2 : Plots of accretion and wind branches for $log(\frac{\lambda}{\lambda_k})$ vs $log(\mathpzc{x})$ for rotating black hole ($spin~ parameter~\mathpzc{J}=0.5$) with different viscosity parameter (showed in different colours). (a) For adiabatic fluid accretion (b) For MCG accretion.}
\end{figure}
\begin{figure}
	\begin{center}
		Fig 2.3(a)\\
	\end{center}
	\includegraphics[height=10cm,width=16cm]{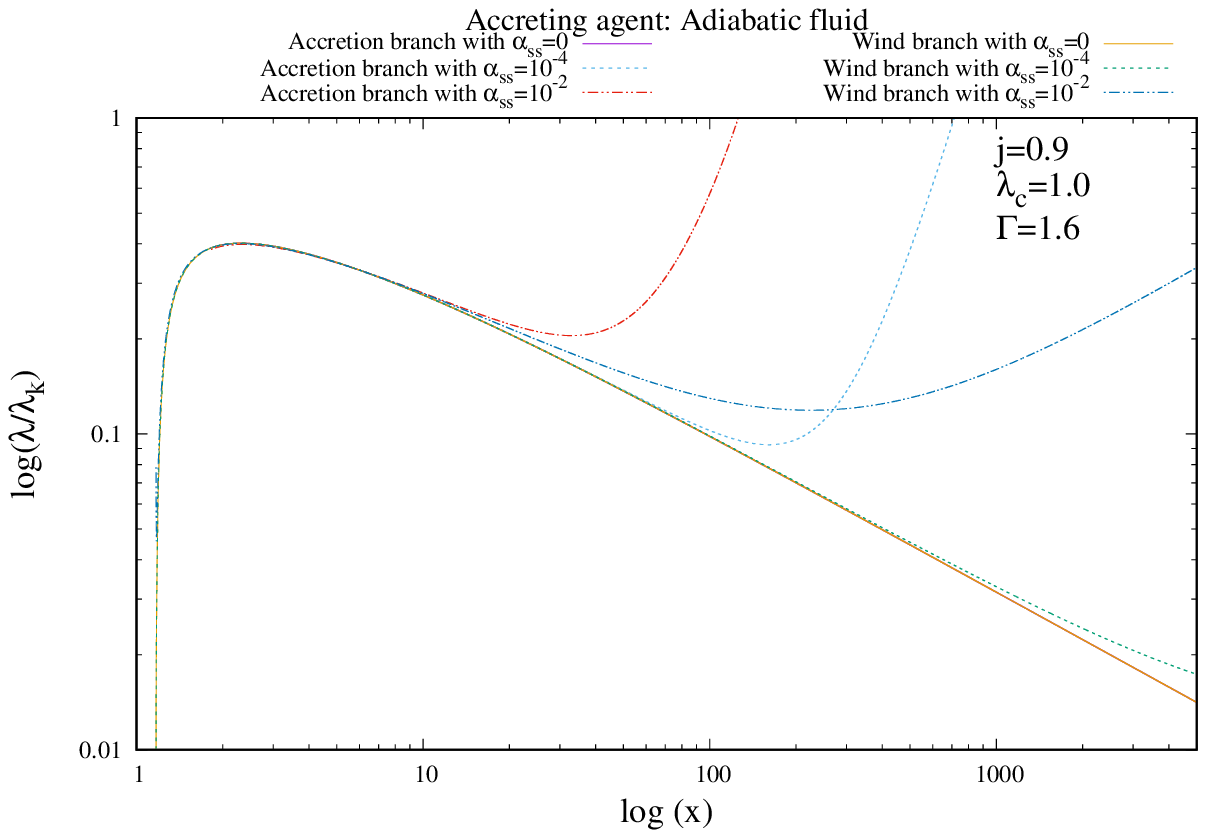}
	\begin{center}
		Fig 2.3(b)\\
	\end{center}
	\includegraphics[height=10cm,width=16cm]{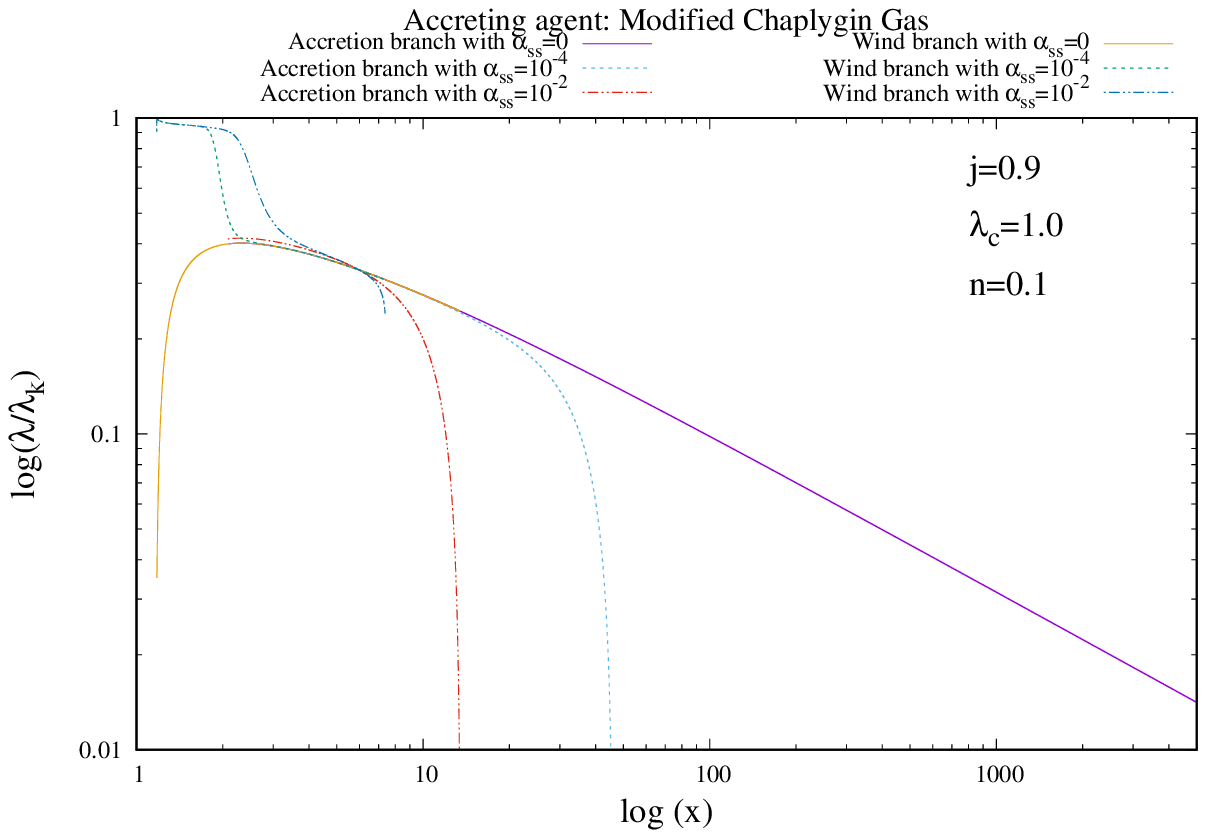}\\
	{Figure 2.3 : Plots of accretion and wind branches for $log(\frac{\lambda}{\lambda_k})$ vs $log(\mathpzc{x})$ for rotating black hole ($spin~ parameter~\mathpzc{J}=0.9$) with different viscosity parameter (showed in different colours). (a) For adiabatic fluid accretion (b) For MCG accretion.}
\end{figure}
\begin{figure}
	\begin{center}
		Fig 3.1(a)\\
	\end{center}
	\includegraphics[height=10cm,width=16cm]{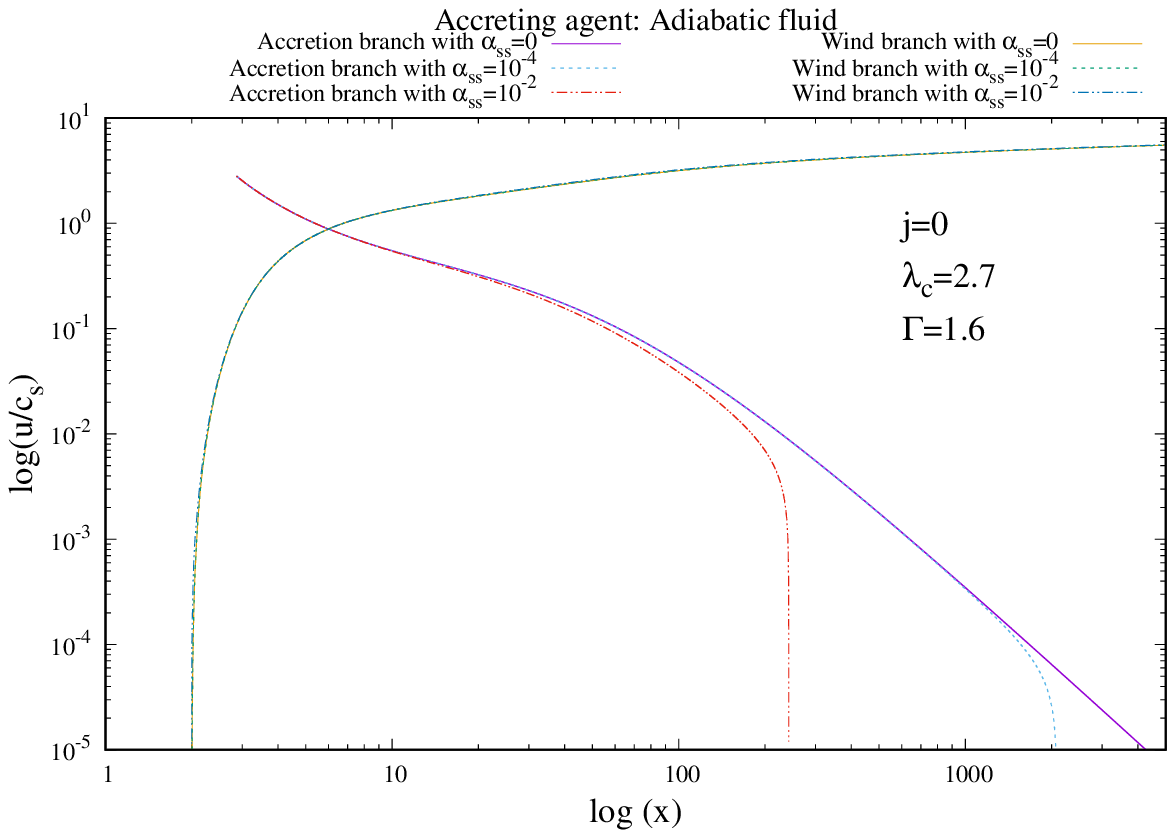}
	\begin{center}
		Fig 3.1(b)\\
	\end{center}
	\includegraphics[height=10cm,width=16cm]{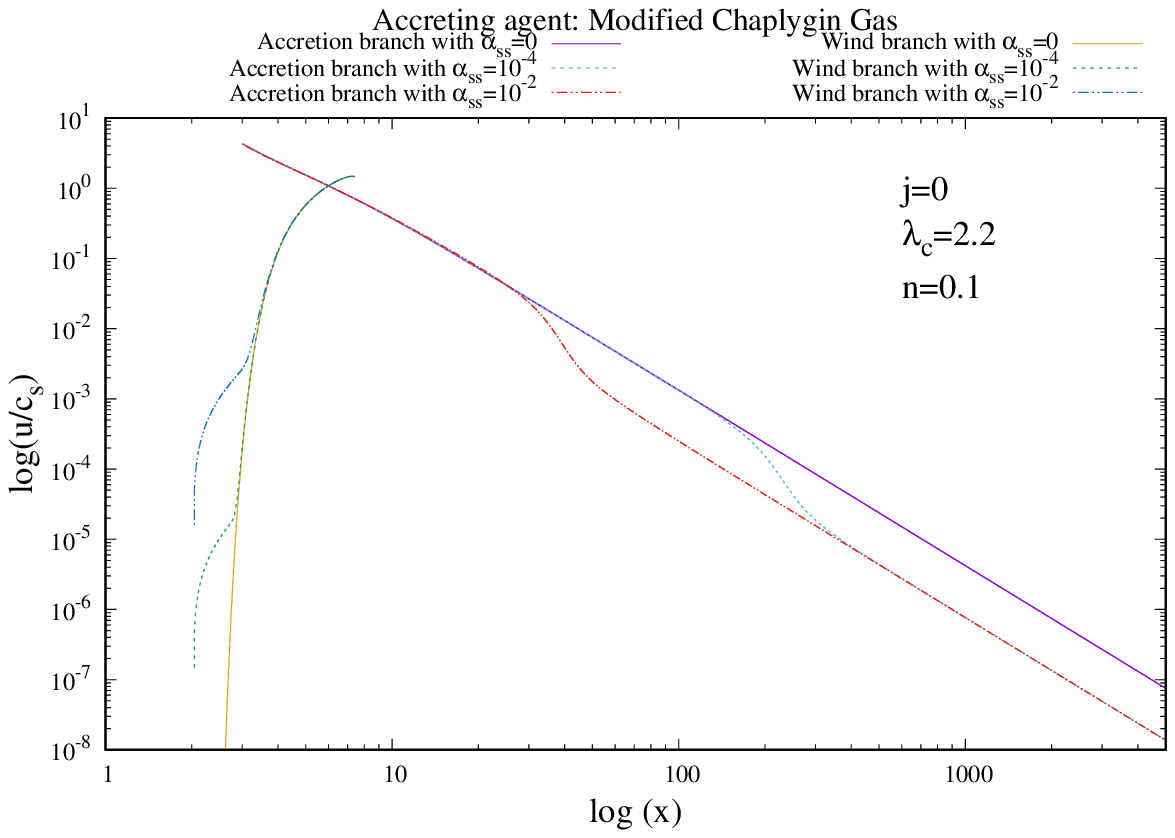}\\
	{Figure 3.1 : Plots of accretion and wind branches for $log(\frac{\mathpzc{U}}{c_s})$ (Mach number) vs $log(\mathpzc{x})$ for non-rotating black hole with different viscosity parameter (showed in different colours). (a) For adiabatic fluid accretion (b) For MCG accretion.}
\end{figure}

\begin{figure}
	\begin{center}
		Fig 3.2(a)\\
	\end{center}
	\includegraphics[height=10cm,width=16cm]{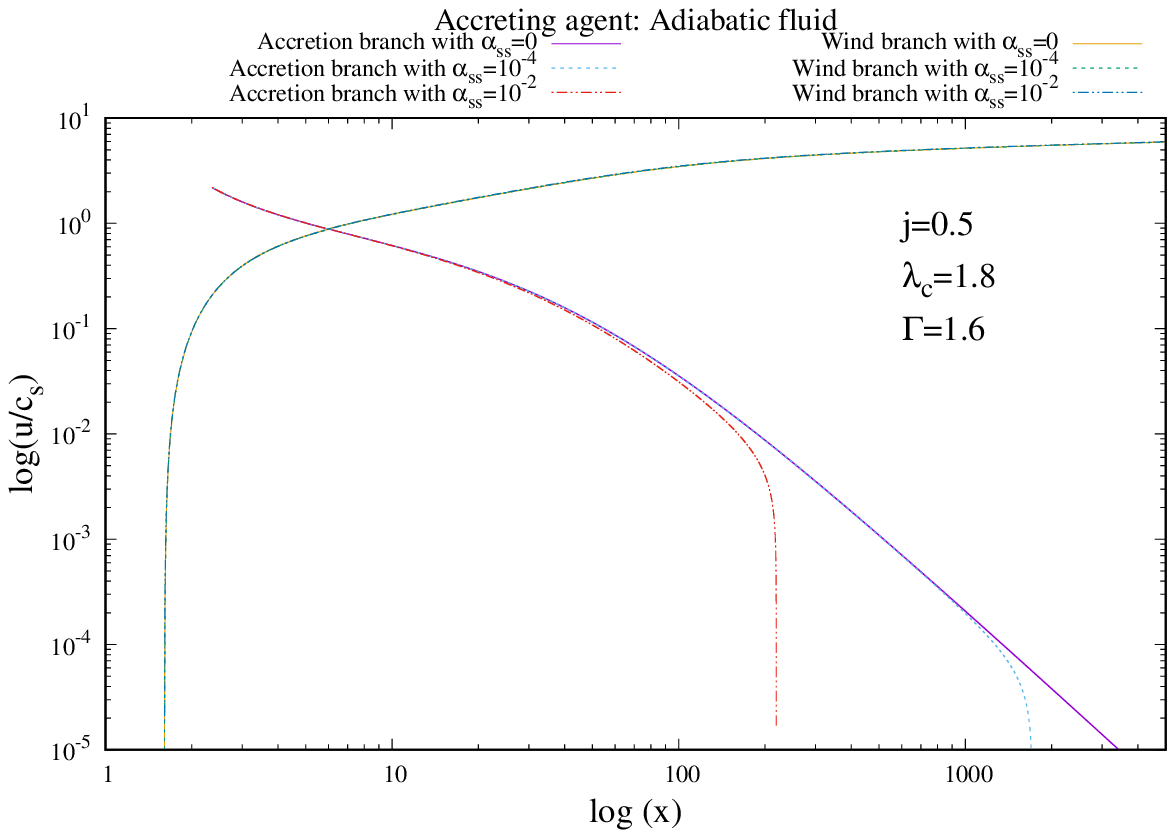}
	\begin{center}
		Fig 3.2(b)\\
	\end{center}
	\includegraphics[height=10cm,width=16cm]{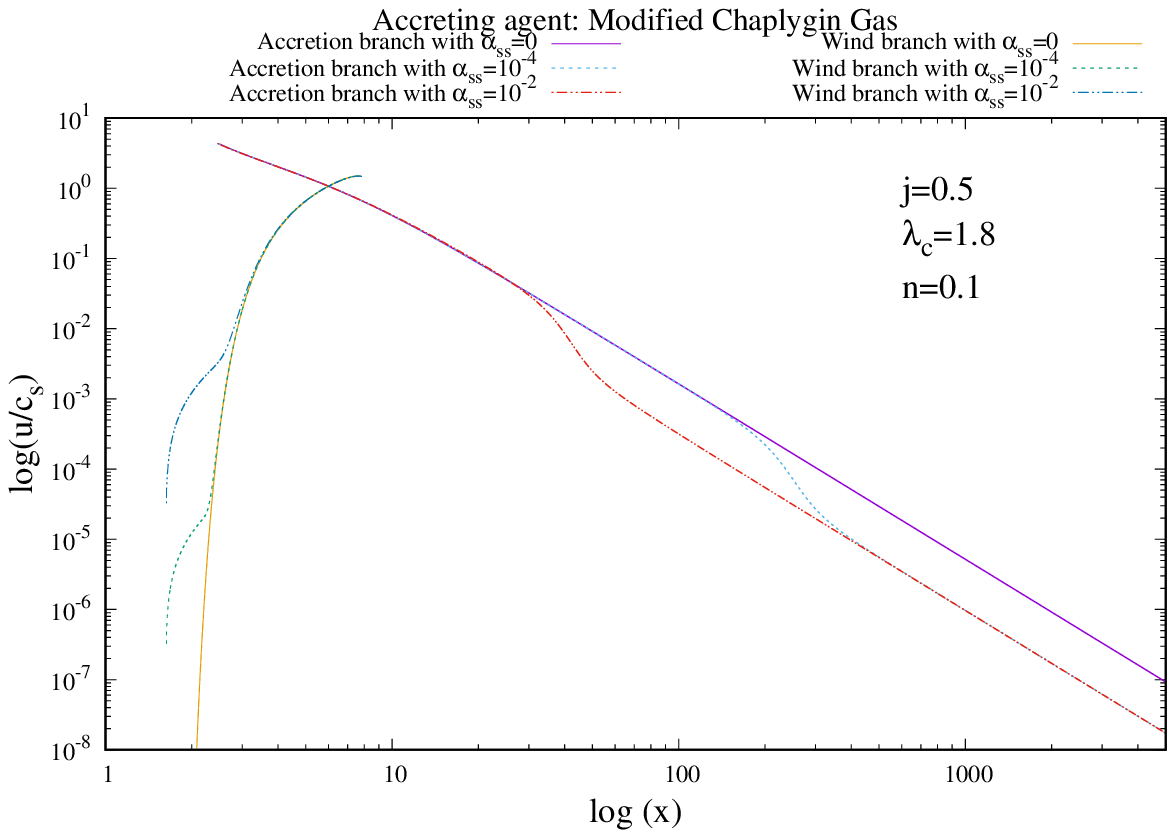}\\
	{Figure 3.2 : Plots of accretion and wind branches for $log(\frac{\mathpzc{U}}{c_s})$ (Mach number) vs $log(\mathpzc{x})$ for rotating black hole ($spin~ parameter~\mathpzc{J}=0.5$) with different viscosity parameter (showed in different colours). (a) For adiabatic fluid accretion (b) For MCG accretion.}
\end{figure}

\begin{figure}
	\begin{center}
		Fig 3.3(a)\\
	\end{center}
	\includegraphics[height=10cm,width=16cm]{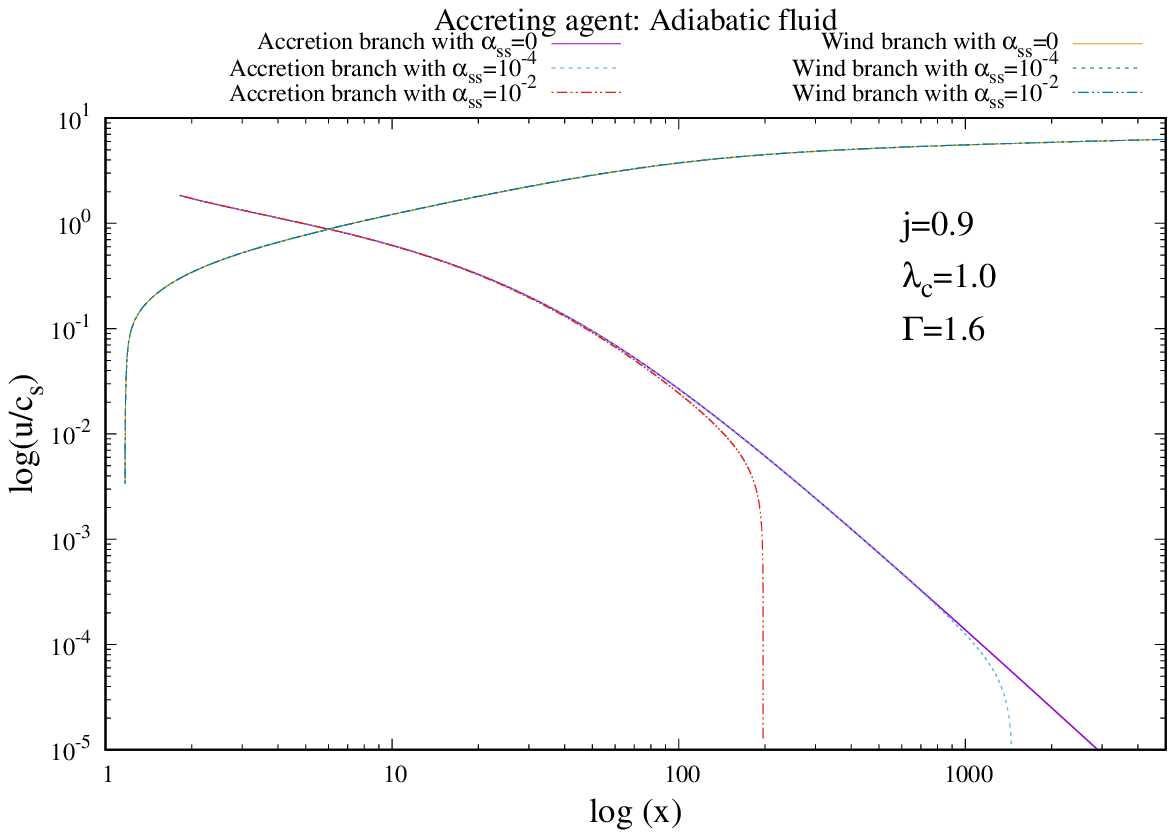}
	\begin{center}
		Fig 3.3(b)\\
	\end{center}
	\includegraphics[height=10cm,width=16cm]{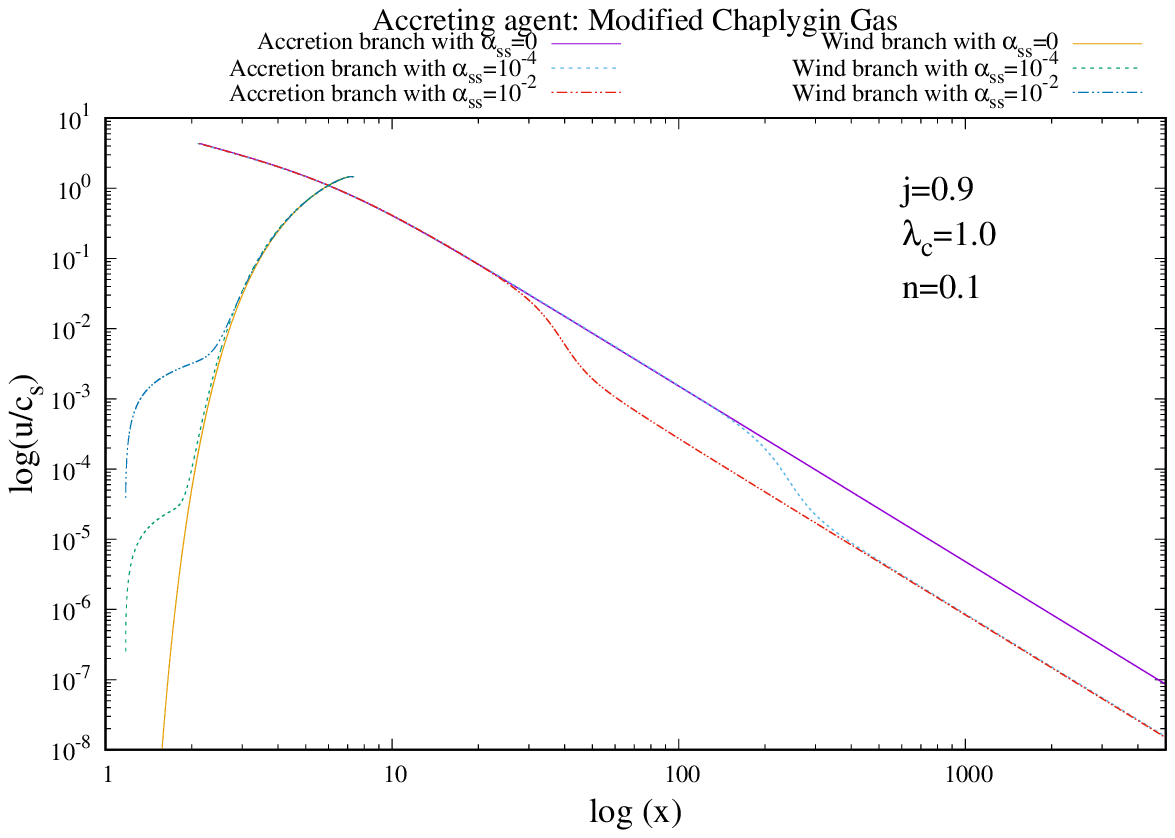}\\
	{Figure 3.3 : Plots of accretion and wind branches for $log(\frac{\mathpzc{U}}{c_s})$ (Mach number) vs $log(\mathpzc{x})$ for rotating black hole ($spin~ parameter~\mathpzc{J}=0.9$) with different viscosity parameter(showed in different colours). (a) For adiabatic fluid accretion (b) For MCG accretion.\\}
\end{figure}

Solving this problem numerically, we plot $log(c_s)$ vs $log(\mathpzc{x})$ curves in figure 1.1(a)-(b), 1.2(a)-(b) \& 1.3(a)-(b).

In fig 1.1(a)-(b), we have plotted $log(c_s)$ vs $log(\mathpzc{x})$ for nonrotating black holes. Among these, 1.1a is for adiabatic accretion and 1.1(b) is for Chaplygin gas as accreting agent. Purple solid, Blue dashed and Red double dot dashed lines are for the accretion. Yellow solid, green dashed and blue double dot dashed lines are for the wind branch.  

First, we will discuss the graphs of adiabatic flow for different viscosity. Very far from the black hole, sonic speed in wind flow is less than that of acrretion flow. As we move towards the black hole, these speeds increase. But as we cross the critical point, sonic speed for wind becomes higher than that of accretion. In inner zone of the disc, sonic speed increases very slowly. Pseudo Newtonian formalism is observed to empower sonic speed to diverge to the speed of light in the near vicinity of Schwarzschild radius. However, in contrast to 3-velocities,  in extreme relativistic regimes, sonic speed is supposed to converge to the upper limit $c/\sqrt{3}$ \cite{Kato:1998}. J. Fukue has proposed a modified sonic speed for Paczinsky Wiita PNF as $c_s^{PN}$ and forced the new sonic speed to converge to the said upper limit as approaching to the Schwarzschild radius \cite{Fukue2004}. Redefining a sonic speed for Mukhopadhyay PNF is kept for a future scope. For now, we consider the form proposed by Fukue and redraw the curves. These are found to stay below the upper limit of $c/{\sqrt{3}}$.

Speed of light in vacuum $c = 299792458~ m/s$. This points that the upper limit of sonic speed should be $c/{\sqrt{3}} = 173085256.3~ m/s$. Now taking log, we get $log (c/{\sqrt{3}})=8.238260076$.
In the figures of $log(c_s)$ vs $log(\mathpzc{U})$, the sound speed for our accretion system lies below the limit of $c/\sqrt{3}$.

Conclusions regarding the temperature of the flow can be drawn from this information. As a flow preceedes towards the central gravitating object, temperature $\sim {c_s}^2$ increases, but slowly. On the other hand, wind, i.e., the flow which is going out of the disc is with very high temperature near the black hole and looses the temperature quickly as it  moves outwards. As we increase the viscosity, the slope of sonic speed increment of both the branches increases towards black hole. This means angular momentum transport is a quick process generated from viscosity and temperature increases rapidly due to this. 

Now, looking towards fig.1.1(b) graphs for Chaplygin gas, we do not exactly see the previous case explained. Both the sonic speed decreases as we go towards the black hole. Sonic speed for accretion decrease slowly whereas wind decreases rapidly. Wind sonic speed is higher than accretion sonic speed if $\mathpzc{x}>{\mathpzc{x}}_c$ and opposite in inner region. 

Physically, this interpretes that the temperature falls as we move towards the central black hole via an accretion branch. Energy to be fallen is radiated away before it falls if the accretion fluid is of Chaplygin gas type. Dark energy wind shows abrupt increment in temperature as it goes off far from the black hole. When dark energy is winded, it carries much energy out of the disc. If viscosity is increased, a threshold drop in accreting sonic speed is observed. It moves nearer with the increment in viscosity.

Fig. 1.2(a) is for sonic speed for a accretion disc where the central black hole is rotating with spin parameter $\mathpzc{J}=0.5$. For same gravitational scenario, dark energy accretion cases are enlisted in fig. 1.2(b).

With nonzero spin of black hole, sonic speed for adiabatic accretion and wind is rapidly increasing than a non rotating black hole as we go towards with black hole. On the other hand, these values decreases rapidly if dark energy accretes in. So fall of energy is higher for a rotating case than a non rotating counterpart. Same tendencies are carried for fig. 1.3(a)-(b) where $\mathpzc{J=0.9}$. It is seen rotation increases the flow of energy towards black hole for adiabatic case and decreases the flow and vacuums it outward as dark energy is accreting. Viscosity helps in this activities. As we move far from the central black hole, pressure increases as density is low for dark energy while enough gravitational pull is not working. As a result, the sonic speed increases suddenly at a larger distances.

To study and understand the disc properties deeply, we will study the ratio of specific angular momentum to Keplerian angular momentum in fig. 2.1(a)-(b)(for nonrotating black holes), fig. 2.2(a)-(b) ($\mathpzc{J}=0.5$), fig. 2.3(a)-(b) ($\mathpzc{J}=0.9$). Non rotating adiabatic black hole accretions shows both accretion and wind's ratio of $\frac{\lambda}{\lambda_k}$ decrease as we move far from the black hole. This supports normal angular momentum transport method. If $\mathpzc{J}=0.5$, accretion $\frac{\lambda}{\lambda_k}$ reduces as $\mathpzc{x}$ increases and after reaching a local minima, it increases and becomes equal to 1 at $\mathpzc{x}=\mathpzc{x_1}$ (say). Here the disc is terminated. In figure 2.1(a), $\mathpzc{J}=0.9$, same pattern is followed and $\frac{\lambda}{\lambda_k}$ turns equal to 1 at $\mathpzc{x}=\mathpzc{x_2}$ (say) where $\mathpzc{x_2} > \mathpzc{x_1}$. We find the reason if spcific angular momentum becomes equal to Keplerian angular momentum, the disc is truncated hereafter. No physical disc is found beyond that. Viscosity increases the accretion rotation as compared to the Keplerian angular momentum in outer disc regions and this causes a centrifugal outward disturbances and thus the disc terminates. Nonrotating nonviscous Chaplygin gas case shows almost the same nature of 2.1a. Difference is noted once viscosity is introduced, the ratio starts to fall in the outward region. More viscosity forces to drop the accretion angular momentum to fall in nearer distance to the black hole. So dark energy opposes the rotation of the disc as we go outwards and hence the disc is truncated. Rotating black holes ($\mathpzc{J}=0.5$ and $0.9$) show almost similar tendencies in fig. 2.2(a)-(b) and fig. 3.2(a)-(b). But rotation drops down the dark energy $\frac{\lambda}{\lambda_k}$ in nearer regions than nonrotating cases. 

Now we will study the variations in Mach number with distance from the black hole in the fig. 3.1(a)-(b), 3.2(a)-(b) and 3.3(a)-(b).

Compressibility of a fluid flow can be best understood by the measure of its Mach number. Mach number's graph are not making after each other like $\frac{\lambda}{\lambda_k}$ curves for nonrotating, nonviscous cases for adiabatic and Chaplygin gas flow. Accretion Mach number increases steeply as we go towards the central black hole. On the other hand wind Mach number increases as we go far from the black hole. Accretion is transonic in inner region whereas wind shows the opposite nature. As the rotation is increased, adiabatic wind turns hypersonic in the outer regions. Besides Chaplygin gas accretion and wind both turn supersonic or hypersonic in the inner disc part. It is to be noted that density changes faster than the change of velocity by a factor of Mach number squared. 

In general, when a system starts to move with hypersonic speed, above shock created by this is followed by  an increment of density. This is again followed by a decrease in volume and gives birth of small shock stand off distance. Changes in entropy across the shock also increases and a highly vertical flow should be observed. Relation between radial inward speed, Mach number and density of the flow can be related as $-(Mach~number)^2 \frac{dv}{v} = \frac{d\rho}{\rho}$. Hence when Mach number is small (in farthest accretion and nearest wind) compressibility can be ignored. On the other hand compressibility withold a good effect if Mach number is high (inner accretion and outer wind).

 Mathematically, the slim disk equations form an eigen value problem, with the eigen value being the angular momentum of matter crossing the black hole horizon. The physical reason for the eigen value nature of the problem is that the black hole accretion must necessarily be transonic, with the sonic point  being the critical (saddle) point of the slim disk differential equations. The regularity conditions at the sonic point assure that its location is very close to ISCO (Innermost Stable Circular Orbit).

To locate the density jumps across the shock, we take the help of the article \cite{Biswas_2019}. At each radius, the value of $\partial_{\phi} \rho({\mathpzc{x}}, \phi)$ can be calculated to identify the region where the value exceeds one. Number of shocks, jump across the shock, azimuthal width of the wake etc can be watched. In this present article we try not to illustrate the issue to avoid unnecessary volumetric increase of the article.
\section{Brief Discussions and Conclusions}
In this article, we study different properties of viscous dark energy accretion. Variations of sonic speed, specific angular momentum to Keplerian angular momentum ratio and Mach number, i.e., fluid flow to sonic speed ratio are plotted and analysed. To set the relativistic attractive force's effects towards the central super massive black hole, we have chosen the pseudo Newtonian force method and precisely a particular pseudo Newtonian potential which at a distance $\mathpzc{x}$ from the center of a black hole spinning with parameter $\mathpzc{J}$, measures the force applicable on a particle. Navier Stokes equations, equation of continuity and equation of state are used to form a system of three simultaneous differential equations to study a thin viscous accretion disk around a rotating super massive black hole. We collect the initial conditions from the continuity and physical properties of the flow in the disc. We study the sonic speed vs radial distance curves in three different sets of two rows of three graphs. Sets are for different spin parameters $\mathpzc{J}=0$, $0.5$ and $0.9$. Rows are for adiabatic and Chaplygin gas cases. Columns are for non viscous, slightly viscous and highly viscous cases. For sonic speed, we observe accretion and wind both sound speeds increase as we go towards the black hole. Wind sonic speed is less than accretion sound speed before the critical point ($\mathpzc{x}>\mathpzc{x_c}$) and opposite if we are inside it ($\mathpzc{x}<\mathpzc{x_c}$). The whole scenario is completely opposite if Chaplygin gas acrretion is taken. As we go towards a black hole the speed decreases in both the branches. Fall in wind branch is steeper as we go towards the black hole. Keeping spin parameter same, if we increase viscosity, we find a fall in accretion branch and this threshold like fall moves towards the black hole as we increase viscosity. Except the fall, there is a general verage tendency of fall in accretion branch sonic speed. This tendency increases as we increase spin. 

We have seen in the works \cite{Dutta2017:SDRBEPJC2,Dutta2019:Dutta,Biswas_2019} that the accretion disc is truncated as Chaplygin gas is accreted. The cause is better understood if we study the $\frac{\lambda}{\lambda_k}$ curves. For adiabatic case this ratio turns equal to $1$ as we go far from the black hole in the wind branch. For Chaplygin gas this ratio falls abruptly. It shows Chaplygin gas resists even the disc to rotate. Resistance increases as we increase spin as well as viscosity. 

Next we move towards the study of Mach number. Mach number is high in inner region of accretion and outer region of wind branch. For adiabatic fluid, wind Mach number is sufficiently high as we go far from the black hole. This signifies a low accretion speed. For dark energy wind, this is not sufficiently high and is truncated at a sufficiently lower value near to $1$. This signifies ocurrance of high speed at outer wind branch for dark energy accretion. Dark energy accretion, as we go nearer to the central gravitating object, shows high Mach number. This indicates a weak inflow indeed. 

{\bf Achkowledgements : } This research is supported by the project grant of Government of West Bengal, Department of Higher Education, Science and Technology and Biotechnology (File no:- $ST/P/S\&T/16G-19/2017$). SD and PB thank to Department of Higher Education and Technology, Government of West Bengal for awarding Swami Vivekananda Merit Cum Means Junior Research Fellowship. RB thanks IUCAA for providing Visiting Associateship. 
\bibliography{paper}

\end{document}